\documentclass[a4paper]{article}

\usepackage{INTERSPEECH_v2}
\usepackage{amsmath,graphicx}
\usepackage{epsfig}
\usepackage{times}
\usepackage{hyperref}
\usepackage{graphicx}
\usepackage{amsmath, amsthm, amssymb}
\usepackage{color}
\usepackage{colortbl}
\usepackage{algorithm}
\usepackage{algorithmic}
\usepackage{subfigure}
\usepackage{subscript}
\usepackage{multirow}
\usepackage{bm}
\usepackage{amsfonts,amssymb}

\title{Multilingual Training and Cross-lingual Adaptation on CTC-based Acoustic Model}
\name{Sibo Tong$^{1,2}$, Philip N. Garner$^1$, Herv{\'e} Bourlard$^{1,2}$}
\address{
  $^1$Idiap Research Institute, Martigny, Switzerland\\
$^2$Ecole Polytechnique F\'{e}d\'{e}rale de Lausanne (EPFL), Switzerland}
 
\email{\{sibo.tong, phil.garner, bourlard\}@idiap.ch}

\begin{document}

\maketitle
\begin{abstract}
 Multilingual models for Automatic Speech Recognition (ASR) are attractive as
  they have been shown to benefit from more training data, and better lend
  themselves to adaptation to under-resourced languages.  However,
  initialisation from monolingual \emph{context-dependent} models leads to an
  explosion of context-dependent states.  Connectionist Temporal Classification
  (CTC) is a potential solution to this as it performs well with monophone
  labels.

  We investigate multilingual CTC in the context of adaptation and
  regularisation techniques that have been shown to be beneficial in more
  conventional contexts.  The multilingual model is trained to model a
  universal International Phonetic Alphabet (IPA)-based phone set using the CTC
  loss function.  Learning Hidden Unit Contribution (LHUC) is investigated to
  perform language adaptive training.  In addition, dropout during cross-lingual adaptation is also studied and tested in order to
  mitigate the overfitting problem.
  
  Experiments show that the performance of the universal phoneme-based CTC
  system can be improved by applying LHUC and it is extensible to
  new phonemes during cross-lingual adaptation.  Updating all the parameters
  shows consistent improvement on limited data.  Applying dropout during
  adaptation can further improve the system and achieve competitive performance
  with Deep Neural Network / Hidden Markov Model (DNN/HMM) systems on limited
  data.
\end{abstract}

\section{Introduction}
\label{sec:intro}
Automatic speech recognition (ASR) systems have improved dramatically in recent years. Although it has been shown that recognition accuracy can reach human parity on certain tasks \cite{xiong2017microsoft}, building ASR systems with good performance requires a lot of training data. While sufficient data is available for languages like English, issues with data scarcity arise for under-resourced languages. Recently there is increased interest in rapidly developing high performance ASR systems with limited data. 

A common solution is to explore universal phonetic structures among different languages by sharing the hidden layers in deep neural networks (DNNs). The hidden layers can be considered to be a universal feature extractor. Therefore, they can be trained jointly using data from multiple languages to benefit each other \cite{huang2013cross,vesely2012language}. Speech recognition systems built with multilingual DNNs have been shown to provide consistent advantages especially for low-resourced languages \cite{huang2013cross, vu2013multilingual,tuske2013investigation,gales2014speech}. Another common approach for creating models for low-resourced languages is to transfer the knowledge learned from other well-resourced languages to the target language. The bottleneck approach extracts language-independent phonetic knowledge from a bottleneck layer of a multilingual model and uses bottleneck features as additional input to train the acoustic model of a target language \cite{thomas2012multilingual, grezl2014adaptation}. Knowledge can also be transferred by replacing the output layer of a well trained model and re-training the model to predict the targets of low-resourced languages \cite{huang2013cross,ghoshal2013multilingual}.

All of these models are based on a conventional DNN/HMM framework \cite{bourlard2012connectionist,hinton2012deep}. In order to perform well, DNNs model context-dependent states to mitigate the error associated with the Markov assumption. However, it creates more challenges for multilingual and cross-lingual ASR because of the large increase in context dependent labels arising from the phone set mismatch.
Although approaches to adapt cluster trees have been proposed \cite{schultz2000polyphone}, the simple and effective way is to replace the whole output layer of a DNN with new targets, or to train a completely new network using bottleneck features. Recently, the Connectionist Temporal Classification (CTC) framework has been successful in ASR \cite{graves2006connectionist}. In CTC training, the neural network is trained to convert a sequence of acoustic features into a sequence of phones or graphemes. CTC based systems learn to model context implicitly by the use of a recurrent neural network (RNN). Even monophone-based CTC systems can achieve equal or better performance than DNN/HMM hybrid systems when a large amount of data is available \cite{sak2015learning,miao2016empirical}. Thus, CTC gets around the problem of context-dependent state mismatch, and does not require prior alignments between the input and output, making the multilingual and cross-lingual modeling simpler and more straightforward.

CTC-based models, however, are more sensitive to the amount of training data. The advantage of CTC training over DNN/HMM can be exploited when adequate data is available. Therefore, we hypothesize that training the CTC model multilingually can further exploit the CTC network by sharing data from multiple languages and that language adaptive training can also boost the performance as in DNN/HMM \cite{tong2017investigation}. To this end, we discuss the universal phoneme-based multilingual CTC model and language adaptive training in Section 2. Given the fact that the multilingual CTC model outputs monophone targets, we hypothesize that the universal phoneme-based multilingual CTC model can serve as a strong prior model when cross-lingual adaptation to a target language is required. Instead of removing the entire output layer and discarding all the information, the output layer of multilingual CTC model can be retained and easily extended to the unseen phonemes in the target languages. Different cross-lingual adaptation approaches based on the CTC framework are discussed in Section 3. In order to minimize the overfitting problem observed in preliminary experiments with CTC, dropout technique is introduced in Section 4. Experimental results and analysis are provided in Section 5. Finally, Section 6 concludes the paper.



\section{Multilingual CTC Neural Network}
\label{sec:format}
\subsection{CTC-based Acoustic Model}
The Connectionist Temporal Classification (CTC) approach
\cite{graves2006connectionist} is an objective function that allows an
end-to-end training without requiring any frame-level alignment between the
input and target labels. CTC allows repetitions of output labels and extends
the set of target labels with an additional \textit{blank} symbol, which
represents the probability of not emitting any labels at a particular time
step. It introduces an intermediate representation called the CTC
\textit{path}. A CTC path is a sequence of labels at the frame level, allowing
repetitions and the blank to be inserted between labels. The label sequence can
be represented by a set of all the possible CTC paths that are mapped to it.

For an input sequence $ \mathbf{X} = (\mathbf{x}_1, \dots , \mathbf{x}_T )$,
the conditional probability $P(\mathbf{y}|\mathbf{X})$ is then obtained by
summing over all the probabilities of all the paths that correspond to the
target label sequence $\mathbf{y}$ after inserting the repetitions of labels
and the blank tokens, i.e.,
\begin{equation}
P(\mathbf{y}|\mathbf{X})=\sum_{\hat{\mathbf{y}}\in \Omega(\mathbf{y})} P(\hat{\mathbf{y}}|\mathbf{X})=\sum_{\hat{\mathbf{y}}\in \Omega(\mathbf{y})}\prod_{t=1}^T P(\hat{y}_t |\mathbf{x}_t)
\label{eq:ctc}
\end{equation}
where $\Omega(\mathbf{y})$ denotes the set of all possible paths that
correspond to $\mathbf{y}$ after repetitions of labels and insertions of the
blank token. The conditional probability of the labels at each time step,
$P(\hat{y}_t|\mathbf{x}_t)$, is estimated using a neural network. The model can
be trained to maximize Equation \ref{eq:ctc} by using gradient descent, where
the required gradients can be computed using the forward-backward algorithm
\cite{graves2006connectionist}.

\subsection{Universal Phone Set Multilingual CTC}
The main goal of multilingual acoustic modelling is to share the acoustic data across multiple languages in order to learn the common properties shared among languages. Many present-day languages evolved from common ancestors. It is therefore natural that they share some common graphemes and phonemes. Very recently, building end-to-end multilingual speech recognition systems by use of a universal grapheme set has been investigated \cite{kim2017towards,toshniwal2017multilingual}. However, graphemes can differ a lot from language to language.  Languages that have nothing in common in terms of graphemes also share some common phonemes. With this motivation, we propose a multilingual architecture that uses a universal output label set consisting of the union of all phonemes from the multiple languages. This universal phone set can be either derived in a data-driven way, or obtained from the International Phonetic Alphabet (IPA). In this study, the monolingual phones are merged if they share the same symbol in the IPA table. The network is trained to model the universal phoneme targets using the CTC loss function on data from multiple languages.

\subsection{Learning Hidden Unit Contribution for Language Adaptive Training}
Since the multilingual CTC network models IPA targets, it may suffer the same problem as the IPA-DNN. Learning Hidden Unit Contribution (LHUC) was first proposed as a method for speaker adaptation \cite{swietojanski2014learning,swietojanski2016sat}. It linearly re-combines hidden units in a speaker- or environment-dependent manner. Further investigation of LHUC in language adaptive training is provided in \cite{tong2017investigation}. Given language-specific data, LHUC re-scales the contributions (amplitudes) of the hidden units in the model without actually modifying their feature receptors. A language-dependent amplitude function is introduced to modify $\mathbf{o}_i^{sl}$ , the hidden unit output of unit $i$ in layer $l$ for language $s$:
\begin{equation}
\label{eq:lhuc}
\mathbf{o}_i^{sl}=\xi(r^{sl}_i) \cdot \psi_i (\mathbf{o}^{l-1})
\end{equation}
$r^{sl}_i  \in \mathbb{R}$  is an adaptable language-dependent parameter, re-parametrised by a function $\xi : \mathbb{R} \rightarrow \mathbb{R}^+$. A sigmoid function with range $(0,2)$ is usually used. $\psi$ is the transformation function in a hidden layer. It can be, for instance, a feedforward or recurrent connection with non-linear activation or a Long Short-Term Memory (LSTM) block. $\psi_i$ is the $i^{th}$ row of the corresponding activations.

The hidden units are trained to capture both good average representations and language-specific representations by estimating language-specific hidden unit amplitudes for each training language. In this paper, LHUC is further combined with the CTC loss function in the context of language adaptive training.
\begin{figure}[!tp]
\centerline{\epsfig{figure=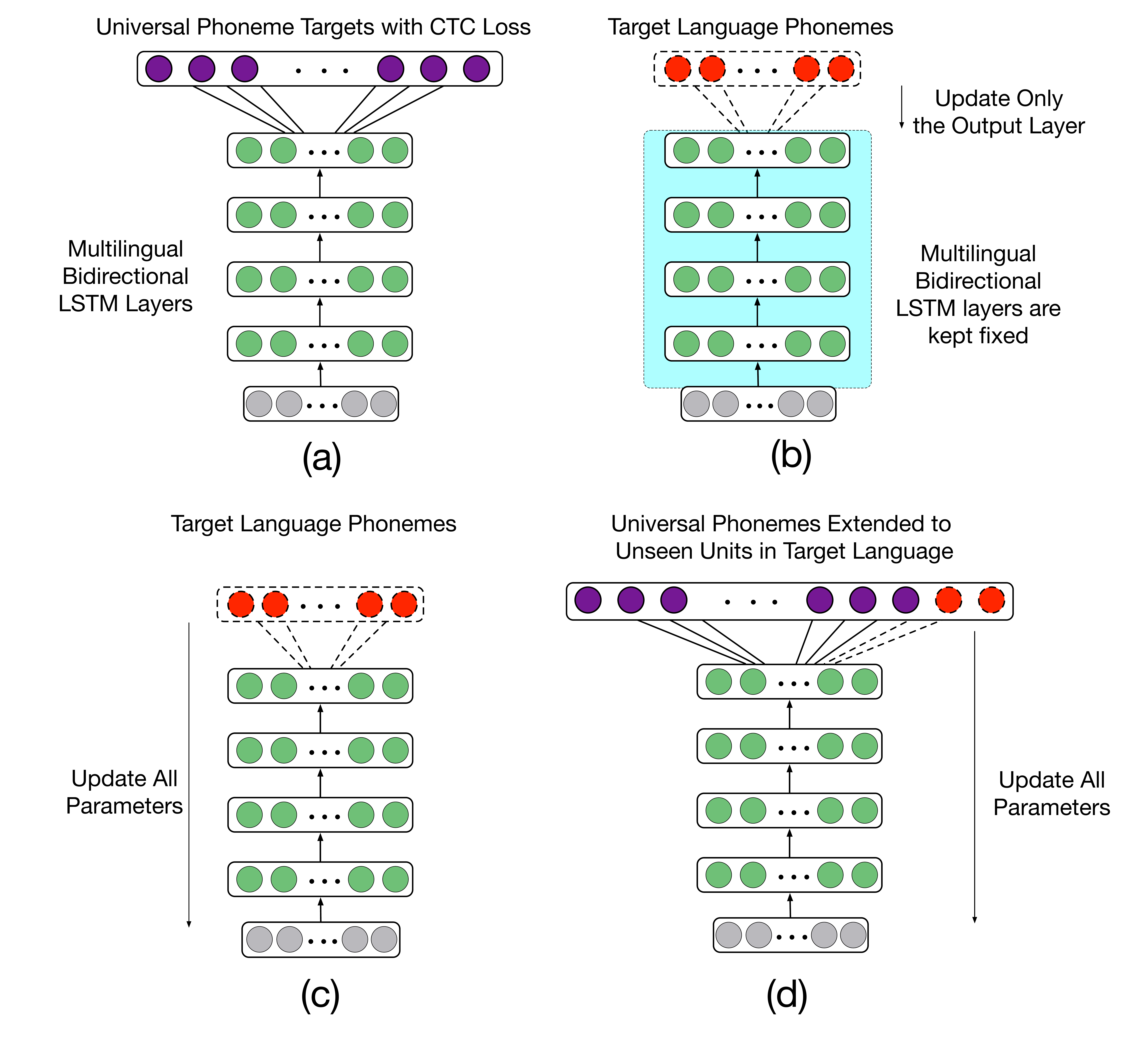,width=100mm}}
\caption{\it Approaches to adapt multilingual CTC model to the target language. (a) shows the multilingual CTC model. In (b), a new output layer replaces the multilingual targets. The hidden layers are fixed and only the output layer is re-estimated.  We can also update all the parameters as shown in (c). In (d), the multilingual CTC model is extended to new phonemes by adding new connections. Adaptation is performed by updating all the parameters. }
\label{fig:crossadpt}
\end{figure}
\section{Cross-lingual Adaptation}
In the DNN framework, the shared hidden layers extracted from the multilingual DNN can be considered to be an intelligent feature extractor and are transferable across languages \cite{huang2013cross}. It is therefore interesting to investigate if the hidden layers in a CTC-based model can be carried over to distinguish phonemes in new languages. 

The basic procedure of cross-lingual model adptation on a CTC model is simple. As first proposed for DNN models \cite{huang2013cross}, the output layer is removed and a new randomly initialized softmax layer, corresponding to the target language phone set, is added on top of the hidden layers. Usually the hidden layers are fixed and only the softmax layer will be re-estimated using training data from the target language. If enough data is available, further tuning of the entire network can be considered.

One major advantage of the universal phoneme-based multilingual CTC model over the multilingual DNN is that mono-phone modeling gets around the problem of mismatch of context-dependent states. It therefore becomes straightforward to extend the existing multilingual model to extra phonemes when a new target language is coming. Therefore, we propose to extend the multilingual output layer by adding connections to the unseen phones of the target language, rather than discarding all the information already learned in the output layer. As is shown in Fig \ref{fig:crossadpt}, those weights connecting to the unseen phones are randomly initialized and trained from scratch. The others can be quickly adapted from the multilingual model with little adaptation data.

\section{Dropout}

In many of our preliminary experiments with CTC, consistent overfitting was observed on limited data. Although adapting from multilingual model mitigates overfitting to some extent, the problem still exists. Dropout has been well established for feed forward networks \cite{srivastava2014dropout}, and it has been also proved to significantly improve the performance of LSTM networks for sequence labelling tasks \cite{reimers2017optimal}. More recently, various approaches of dropout on feedforward and recurrent connections were explored in the context of CTC \cite{billa2017improving}. Inspired by this work, we propose to combine dropout and cross-lingual adaptation to minimize overfitting on low-resourced languages. The dropout approach applied in this work is a combination of dropout on two different levels, as described in \cite{billa2017improving}
\begin{itemize}
\item \textbf{Dropout on feed forward connections} Dropout is applied on the feed forward connections at sequence level where the composite LSTM cell is the unit to be dropped. The dropout mask is retained across a complete utterance to eliminate cross-sampling noise.
\item \textbf{Dropout on recurrent connections} Recurrent dropout without memory loss \cite{semeniuta2016recurrent} is applied to the incremental LSTM cell memory update at sequence level following
\begin{equation}
\mathbf{c}_t=\mathbf{f}_t\odot\mathbf{c}_{t-1}+\mathbf{m}_t\odot\mathbf{i}_t\odot\phi(\mathbf{W}_c\mathbf{x}_t+\mathbf{R}_c\mathbf{h}_{t-1}+\mathbf{b}_c)
\end{equation}
where $\mathbf{c}_t$ is the LSTM cell state at time $t$, $\mathbf{f}_t$ and $\mathbf{i}_t$ respectively denote the forget gate and input gate, $\mathbf{x}_t$ is the input vector at time $t$, $\mathbf{h}_{t-1}$ represents the LSTM output at time $t-1$, $\mathbf{W}_c$, $\mathbf{R}_c$ and $\mathbf{b}_c$ are the corresponding weights and bias, $\mathbf{m}_t$ represents the dropout mask at time $t$. The mask is again retained across a complete sequence.
\end{itemize}
For each minibatch, either a forward or recurrent dropout is picked randomly with equal probability. The combination was observed to outperform single dropout training.

\section{Experiments}
\subsection{Experimental Database}

We investigated the performance of the proposed universal phoneme-based CTC model on English (EN), French (FR), and German (GE). The English data was obtained from the Wall Street Journal (WSJ) corpus \cite{paul1992design}. Data preparation gave us 81 hours of transcribed speech. The French data was extracted from the BREF \cite{lamel1991bref} and GlobalPhone corpora \cite{schultz2013globalphone}, which consist of 120 hours of data. From the German Broadcast News (BCN) corpus \cite{weninger2014broadcast}, we used 136 hours of data for training. In total, 337 hours of multilingual data was used for multilingual CTC training. All the training data is quite clean read speech from similar acoustic conditions. In cross-lingual adaptation experiments, Portuguese and Spanish from GlobalPhone were considered as target low-resourced languages, which have only 21 hours and 16 hours data respectively.

\subsection{Setup}
We used 40-dimensional log-mel filterbank coefficients as acoustic features together with their first and second-order derivatives, derived from 25 ms frames with a 10 ms frame shift. The features were normalized via mean subtraction and variance normalization on a speaker basis. All the monolingual phones were mapped to IPA symbols and we merged the phonemes from EN, FR and GE to create the universal phone set for multilingual training.

The multilingual CTC model has 4 layers of Bidirectional Long Short-Term Memory (BLSTM), with 320 cells in each layer and direction. All the weights in the models were randomly initialized and were trained using stochastic gradient descent with momentum. A learning rate of 0.0004 was used and early stopping on the validation set was applied to select the best model. For decoding, individual weighted finite-state transducer (WFST) decoding graphs were built using language-specific lexica and language models.

%
%
%
%

\subsection{Results}

\begin{table} [t]

\caption{\label{tab:baseline} {\it Comparison between monolingual baseline systems and multilingual training in WER(\%).}}
\centerline{
\begin{tabular}{|c||c c c|}
\hline
system & EN & FR & GE \\
\hline\hline
monolingual CTC & 8.7 & 8.5 & 8.9 \\
universal ML-CTC &  9.0 & 8.1 & 9.0 \\
+LHUC & \bf{8.5} & \bf{7.7} & \bf{8.4} \\
\hline
\end{tabular}}
\end{table}
\subsubsection{Multilingual Training}

This section presents all the experimental results of our study. Previous research has shown that an adequate amount of data is the key to training a good CTC-based system. We first evaluated if a better model can be trained using data from multiple languages. The comparison between multilingual CTC and baseline monolingual CTC systems is listed in Table \ref{tab:baseline}. It shows that monolingual CTC systems still perform better than the multilingual model, even though they were trained only on around 100 hours data. We observed a similar result in our previous work on an IPA-based universal DNN system. Although the universal multilingual modelling enjoys richer data resources, the mixture of data creates more variation, especially for those identical IPA symbols shared among different languages. The result is consistent with another recent independent study \cite{muller2017multilingual}. This motivates us to apply language adaptive training in the multilingual CTC model. As shown in the last row of the table, multilingual CTC combined with LHUC improves the performance and yields better or similar word error rate (WER) to the monolingual CTC in all languages. In our initial experiments, we observed that the multilingual model trained with LHUC cannot yield more improvement over the standard multilingual model when adapted to a new language. Therefore, the standard multilingual model trained on the 3 languages was used as the seed model for the following cross-lingual experiments and is denoted as ML-3.

%
\subsubsection{Cross-lingual Adaptation to Portuguese}
\begin{figure}[htb]
  \centerline{\includegraphics[width=0.5\textwidth]{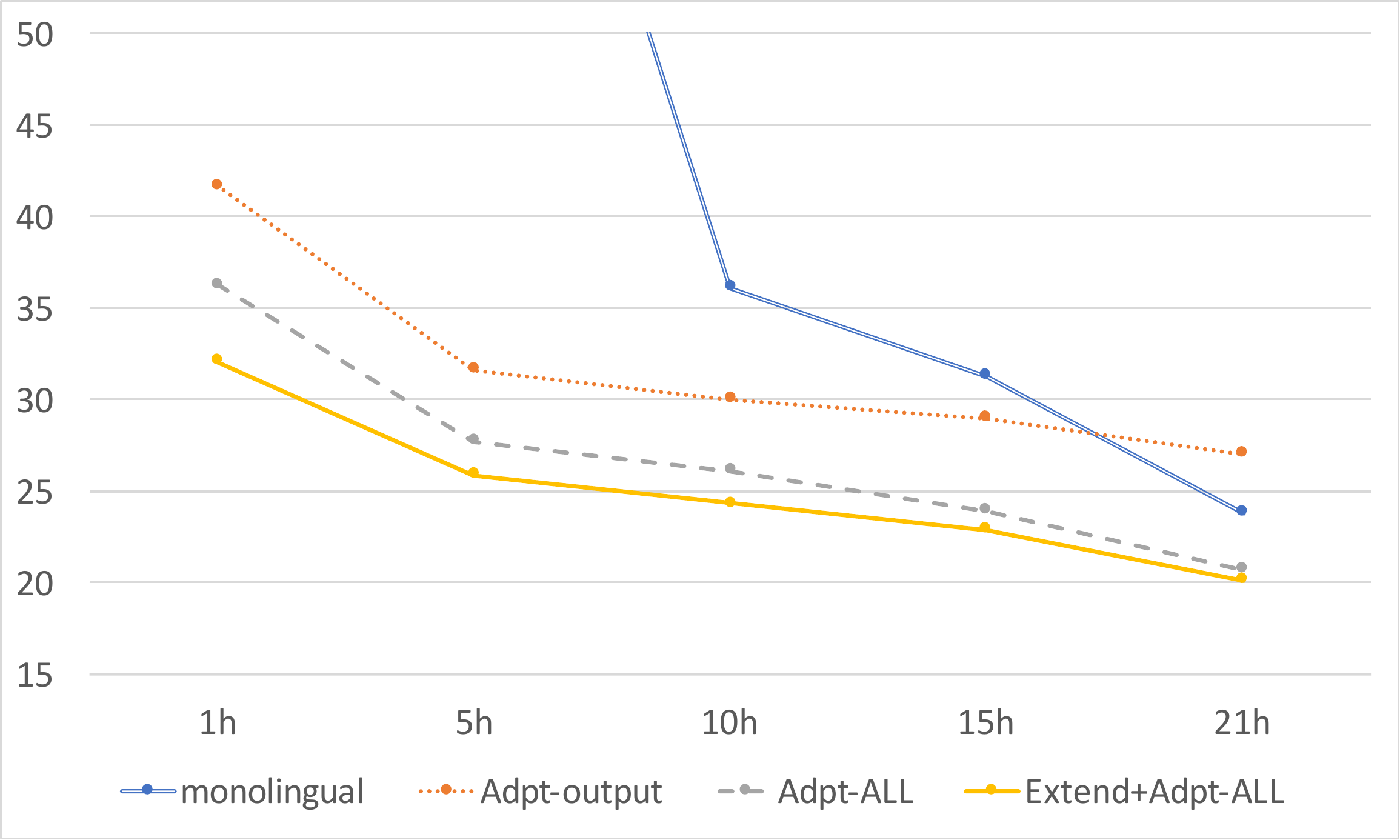}}
\caption{{\it WERs (\%) of different cross-lingual adaptation approaches. The WERs of monolingual CTC models on 1 hour and 5 hours data are above 50\% and exceed the graph region.}}

\label{fig:crossdata}
\end{figure}

While the first goal of this work was to create a universal phoneme based multilingual model, we were interested in its transfer ability to other languages when the training data is limited. In order to perform the cross-lingual adaptation, three approaches were investigated: re-training a new output layer while keeping other parameters fixed; re-training a new output layer and also updating other parameters; and extending the multilingual model by randomly initializing weights between the last layer and the new phonemes and then updating the whole network. Experiments on different amounts of data were conducted using these approaches. Fig. \ref{fig:crossdata} shows all the comparisons.

From the figure, it can be found that all adaptation approaches outperform monolingual CTC training on limited data (less than 15 hours). It is impossible to train a good CTC model using less than 10 hours of data. However, adaptation from the multilingual model can still achieve good performance. When the adaptation data is more than 15 hours, monolingual training beats the adaptation on only the output layer. Moreover, updating all the parameters performs better than only re-training the output layer in all cases, which means the hidden BLSTM layers are not completely transferable like those of the DNN \cite{huang2013cross}. Keeping the multilingual output layer and extending the multilingual network yields additional improvement. However, the difference becomes marginal with the increase of the adaptation data. 

\subsubsection{Cross-lingual Adaptation to Spanish}
Given the above observations, it seems reasonable to hypothesize the coverage of phonemes can affect the performance of cross-lingual adaptation, especially on limited data. To validate our hypothesis, two multilingual models were chosen as seed models for the adaptation to Spanish, namely the model trained on English, French and German (ML-3) and the model adapted from ML-3 to Portuguese (denoted as ML-4). The Spanish phone set consists of 40 phonemes. While 23 of them have been seen in ML-3, ML-4 covers 7 more phonemes because of the presence of Portuguese. Adaptations were conducted based on these 2 seed models. As shown in Table \ref{tab:sp}, adaptation from ML-4 yields consistent improvement over that from ML-3. The multilingual model covers more phonemes after each adaptation. We believe it will become a stronger and stronger multilingual prior model for cross-lingual adaptation by extending the model to more and more languages.
\begin{table} [t]
\caption{\label{tab:sp} {\it Comparison of cross-lingual adaptation from multilingual model to Spanish with different phoneme coverage on different amount of data in WER(\%).}}
\centerline{
\begin{tabular}{c c c c c c}
\hline
Adaptation from & \# Covered phn & 1h & 5h & 10h & All \\
\hline\hline
ML-3 & 23 & 15.4 & 12.7 & 11.3 & 10.6 \\
ML-4 & 30 & 15.0 & 12.2 & 11.2 & 10.4 \\
\hline
\end{tabular}}
\end{table}

\subsubsection{Dropout}
In order to validate our hypothesis that dropout can further minimize overfitting on low-resourced languages, another set of experiments were conducted and compared with the DNN/HMM system. Table \ref{tab:dropout} details our results.

All the systems are trained or adapted on 21 hours of Portuguese data. The DNN/HMM system models 3100 context-dependent states obtained from decision tree clustering and has 6 hidden layers, each consisting of 1024 units. Thus, it contains slightly more parameters (8.8 vs 8.5 million) than the CTC models. It is clear from Table \ref{tab:dropout} that the DNN/HMM performs much better than the CTC model trained from scratch on 21 hours data. Dropout on the monolingual CTC model improves the performance and cross-lingual adaptation from the multilingual CTC yields more improvement. By applying dropout during cross-lingual adaptation, the WER is further reduced. However, the improvement is not as much as that for dropout on monolingual CTC. One conjecture is that both dropout and cross-lingual adaptation model help avoid overfitting on limited data. However, cross-lingual adaptation, with the help of dropout, can achieve competitive performance with the monolingual DNN/HMM system on only 20 hours data; previous research shows at least 100 hours of data is required to enable CTC to reach similar performance \cite{miao2016empirical}.

\begin{table} [t]

\caption{\label{tab:dropout} {\it Comparison between monolingual systems and cross-lingual adaptation combined with dropout on Portuguese in WER(\%).}}
\centerline{
\begin{tabular}{c c }
\hline
system & WER (\%) \\
\hline\hline
monolingual DNN/HMM & 19.5 \\
monolingual CTC & 23.8  \\
+dropout & 21.1 \\
Adapted from ML-3 &  20.5  \\
+dropout & \bf{19.0}  \\
\hline
\end{tabular}}
\end{table}

\section{Acknowledgement}
This work has been conducted with the support of the European Community H2020 “Research and Innovation Action”funding, under “Scalable Understanding of Multilingual Media” (SUMMA) project n. 688139. 
\section{Conclusions}
\label{sec:refs}

It was demonstrated that LHUC can be applied in CTC-based multilingual training and mitigate the problem arising from the mixture of data from various languages. The universal phoneme-based multilingual CTC is extensible to new phonemes during cross-lingual adaptation, and updating all the parameters shows consistent improvement on limited data. Combined with dropout during adaptation, the CTC-based model shows competitive performance with DNN/HMM even on 21 hours data.

\bibliographystyle{IEEEtran}

\bibliography{mybib}

\begin{thebibliography}{10}
\providecommand{\url}[1]{#1}
\csname url@samestyle\endcsname
\providecommand{\newblock}{\relax}
\providecommand{\bibinfo}[2]{#2}
\providecommand{\BIBentrySTDinterwordspacing}{\spaceskip=0pt\relax}
\providecommand{\BIBentryALTinterwordstretchfactor}{4}
\providecommand{\BIBentryALTinterwordspacing}{\spaceskip=\fontdimen2\font plus
\BIBentryALTinterwordstretchfactor\fontdimen3\font minus
  \fontdimen4\font\relax}
\providecommand{\BIBforeignlanguage}[2]{{%
\expandafter\ifx\csname l@#1\endcsname\relax
\typeout{** WARNING: IEEEtran.bst: No hyphenation pattern has been}%
\typeout{** loaded for the language `#1'. Using the pattern for}%
\typeout{** the default language instead.}%
\else
\language=\csname l@#1\endcsname
\fi
#2}}
\providecommand{\BIBdecl}{\relax}
\BIBdecl

\bibitem{xiong2017microsoft}
W.~Xiong, J.~Droppo, X.~Huang, F.~Seide, M.~Seltzer, A.~Stolcke, D.~Yu, and
  G.~Zweig, ``The {Microsoft} 2016 conversational speech recognition system,''
  in \emph{Proceedings of the {IEEE} International Conference on Acoustics,
  Speech and Signal Processing}, 2017.

\bibitem{huang2013cross}
J.-T. Huang, J.~Li, D.~Yu, L.~Deng, and Y.~Gong, ``Cross-language knowledge
  transfer using multilingual deep neural network with shared hidden layers,''
  in \emph{Proceedings of the {IEEE} International Conference on Acoustics,
  Speech and Signal Processing}, 2013.

\bibitem{vesely2012language}
K.~Vesel{\`y}, M.~Karafi{\'a}t, F.~Gr{\'e}zl, M.~Janda, and E.~Egorova, ``The
  language-independent bottleneck features,'' in \emph{Proceedings of the
  {IEEE} Workshop on Spoken Language Technology}, 2012.

\bibitem{vu2013multilingual}
N.~T. Vu and T.~Schultz, ``Multilingual multilayer perceptron for rapid
  language adaptation between and across language families,'' in
  \emph{Proceedings of Interspeech}, 2013.

\bibitem{tuske2013investigation}
Z.~T{\"u}ske, J.~Pinto, D.~Willett, and R.~Schl{\"u}ter, ``Investigation on
  cross-and multilingual {MLP} features under matched and mismatched acoustical
  conditions,'' in \emph{Proceedings of the {IEEE} International Conference on
  Acoustics, Speech and Signal Processing}, 2013.

\bibitem{gales2014speech}
M.~J.~F. Gales, K.~M. Knill, A.~Ragni, and S.~P. Rath, ``Speech recognition and
  keyword spotting for low resource languages: Babel project research at
  {CUED},'' \emph{Spoken Language Technologies for Under-Resourced Languages},
  2014.

\bibitem{thomas2012multilingual}
S.~Thomas, S.~Ganapathy, and H.~Hermansky, ``Multilingual {MLP} features for
  low-resource {LVCSR} systems,'' in \emph{Proceedings of the {IEEE}
  International Conference on Acoustics, Speech and Signal Processing}, 2012.

\bibitem{grezl2014adaptation}
F.~Gr{\'e}zl, M.~Karafi{\'a}t, and K.~Vesel{\`y}, ``Adaptation of multilingual
  stacked bottle-neck neural network structure for new language,'' in
  \emph{Proceedings of the {IEEE} International Conference on Acoustics, Speech
  and Signal Processing}, 2014.

\bibitem{ghoshal2013multilingual}
A.~Ghoshal, P.~Swietojanski, and S.~Renals, ``Multilingual training of deep
  neural networks,'' in \emph{Proceedings of the {IEEE} International
  Conference on Acoustics, Speech and Signal Processing}, 2013.

\bibitem{bourlard2012connectionist}
H.~A. Bourlard and N.~Morgan, \emph{Connectionist speech recognition: a hybrid
  approach}, 2012.

\bibitem{hinton2012deep}
G.~Hinton, L.~Deng, D.~Yu, G.~E. Dahl, A.-r. Mohamed, N.~Jaitly, A.~Senior,
  V.~Vanhoucke, P.~Nguyen, T.~N. Sainath \emph{et~al.}, ``Deep neural networks
  for acoustic modeling in speech recognition: The shared views of four
  research groups,'' \emph{IEEE Signal Processing Magazine}, 2012.

\bibitem{schultz2000polyphone}
T.~Schultz and A.~Waibel, ``Polyphone decision tree specialization for language
  adaptation,'' in \emph{Proceedings of the {IEEE} International Conference on
  Acoustics, Speech and Signal Processing}, 2000.

\bibitem{graves2006connectionist}
A.~Graves, S.~Fern{\'a}ndez, F.~Gomez, and J.~Schmidhuber, ``Connectionist
  temporal classification: labelling unsegmented sequence data with recurrent
  neural networks,'' in \emph{Proceedings of the 23rd international conference
  on Machine learning}, 2006.

\bibitem{sak2015learning}
H.~Sak, A.~Senior, K.~Rao, O.~Irsoy, A.~Graves, F.~Beaufays, and J.~Schalkwyk,
  ``Learning acoustic frame labeling for speech recognition with recurrent
  neural networks,'' in \emph{Proceedings of the {IEEE} International
  Conference on Acoustics, Speech and Signal Processing}, 2015.

\bibitem{miao2016empirical}
Y.~Miao, M.~Gowayyed, X.~Na, T.~Ko, F.~Metze, and A.~Waibel, ``An empirical
  exploration of {CTC} acoustic models,'' in \emph{Proceedings of the {IEEE}
  International Conference on Acoustics, Speech and Signal Processing}, 2016.

\bibitem{tong2017investigation}
S.~Tong, P.~N. Garner, and H.~Bourlard, ``An investigation of deep neural
  networks for multilingual speech recognition training and adaptation,'' in
  \emph{Proceedings of Interspeech}, 2017.

\bibitem{kim2017towards}
S.~Kim and M.~L. Seltzer, ``Towards language-universal end-to-end speech
  recognition,'' \emph{arXiv preprint arXiv:1711.02207}, 2017.

\bibitem{toshniwal2017multilingual}
S.~Toshniwal, T.~N. Sainath, R.~J. Weiss, B.~Li, P.~Moreno, E.~Weinstein, and
  K.~Rao, ``Multilingual speech recognition with a single end-to-end model,''
  \emph{arXiv preprint arXiv:1711.01694}, 2017.

\bibitem{swietojanski2014learning}
P.~Swietojanski and S.~Renals, ``Learning hidden unit contributions for
  unsupervised speaker adaptation of neural network acoustic models,'' in
  \emph{Proceedings of the {IEEE} Workshop on Spoken Language Technology},
  2014.

\bibitem{swietojanski2016sat}
------, ``{SAT-LHUC}: Speaker adaptive training for learning hidden unit
  contributions,'' in \emph{Proceedings of the {IEEE} International Conference
  on Acoustics, Speech and Signal Processing}, 2016.

\bibitem{srivastava2014dropout}
N.~Srivastava, G.~E. Hinton, A.~Krizhevsky, I.~Sutskever, and R.~Salakhutdinov,
  ``Dropout: a simple way to prevent neural networks from overfitting.''
  \emph{Journal of machine learning research}, 2014.

\bibitem{reimers2017optimal}
N.~Reimers and I.~Gurevych, ``Optimal hyperparameters for deep {LSTM}-networks
  for sequence labeling tasks,'' \emph{arXiv preprint arXiv:1707.06799}, 2017.

\bibitem{billa2017improving}
J.~Billa, ``Improving {LSTM-CTC} based {ASR} performance in domains with
  limited training data,'' \emph{arXiv preprint arXiv:1707.00722}, 2017.

\bibitem{semeniuta2016recurrent}
S.~Semeniuta, A.~Severyn, and E.~Barth, ``Recurrent dropout without memory
  loss,'' 2016.

\bibitem{paul1992design}
D.~B. Paul and J.~M. Baker, ``The design for the {Wall} {Street}
  {Journal}-based {CSR} corpus,'' in \emph{Proceedings of the workshop on
  Speech and Natural Language}, 1992.

\bibitem{lamel1991bref}
L.~F. Lamel, J.-L. Gauvain, M.~Esk{\'e}nazi \emph{et~al.}, ``{BREF}, a large
  vocabulary spoken corpus for french1,'' 1991.

\bibitem{schultz2013globalphone}
T.~Schultz, N.~T. Vu, and T.~Schlippe, ``{GlobalPhone}: A multilingual text \&
  speech database in 20 languages,'' in \emph{Proceedings of the {IEEE}
  International Conference on Acoustics, Speech and Signal Processing}, 2013.

\bibitem{weninger2014broadcast}
F.~Weninger, B.~Schuller, F.~Eyben, M.~W{\"o}llmer, and G.~Rigoll, ``A
  broadcast news corpus for evaluation and tuning of {German} {LVCSR}
  systems,'' \emph{arXiv preprint arXiv:1412.4616}, 2014.

\bibitem{muller2017multilingual}
M.~M{\"u}ller, S.~St{\"u}ker, and A.~Waibel, ``Multilingual adaptation of {RNN}
  based {ASR} systems,'' \emph{arXiv preprint arXiv:1711.04569}, 2017.

\end{thebibliography}


\end{document}